\documentclass[referee]{iopart}        
\usepackage{graphicx}
\usepackage{natbib}

\begin{document}
\title[Shallow decay of X-ray afterglow]{Shallow decay phase of the early X-ray afterglow from the external shock in wind environment}

\author{LEI Hai-Dong$^1$, WANG Jiu-Zhou$^2$, LV Jing$^2$ and ZOU Yuan-Chuan$^2$}

\address{$^1$Department of Physics and Electronics, Jianghan University, Wuhan 430056, China; {\it m7205@tom.com} \\ $^2$School of physics, Huazhong University of Science and Technology, Wuhan 430074, China; {\it wjzhust@foxmail.com, lvjing310@smail.hust.edu.cn and zouyc@hust.edu.cn}}

\begin{abstract}
The shallow decay phase of the early X-ray afterglow in gamma-ray bursts discovered by {\it Swift} is a widely discussed topic. As the spectral index does not change at the transferring of the shallow decay phase to a normal phase, it implies this transferring should be a dynamical change rather than a spectral evolution. We suggest both the shallow decay phase and the normal phase are from the external shock in a wind environment, while the transferring time is the deceleration time. We apply this model to GRBs 050319 and 081008, and find them can be well explained by choosing a proper set of parameters.
\end{abstract}

%\maketitle

\section{Introduction}        
\label{intro}

In the {\it Swift} era, thanks to the rapid response, many early X-ray emissions  in gamma-ray bursts (GRBs) were observed (e.g. \citealt{nousek06,obrien06}). Two main unexpected phenomena were found in most bursts: X-ray flares following the prompt burst phase, and a shallow decay phase (\citealt{zhang05,nousek06,obrien06}), which may overlap the X-ray flares. Many theoretical models were proposed to explain the new phenomena (e.g. \citealt{zhang05, toma06, wu06, zou06a, zou06b}). For the shallow decay phase, \citet{nousek06} and \citet{zhang05} supposed it is caused from the late energy re-injection, \citet{toma06} proposed a special jet profile, \citet{zou06a} and \citet{xu10} thought they may be tail emission from a slow ring-like jet, \citet{fan06} and \citet{kong10} considered the possibility by tuning the microphysical parameters, \citet{toma06} suggested summation of many different directed mini-jets may produce the shallow decay radiation, \citet{gs07} suggested a late internal shock emission model, and \citet{yamazaki09} and \citet{liang09} suggested that by shifting the starting time to be $\sim 10^3$s ahead, the shallow decay becomes normal decay automatically. Here we propose an alternative explanation: the external shock (in wind environment) before the deceleration time produces the shallow decay phase of early X-ray emission.

The main features of the shallow decay phase are: the temporal profile is $t^{-\alpha}$ (with $\alpha \sim (0,0.5)$), the spectrum is $\sim \nu^{-1}$, and the spectral index does not change after transferring to the normal decay phase (\citealt{zhang05}). As the temporal and spectral indices are the same at both the shallow decay phase and the normal decay phase, this strongly suggests the emission is from the same region but different dynamics. We check the possibility for the reverse-forward shock model, as for the forward shock and the late external shock, they are from the same region but different dynamics, and find that, in wind environment, the temporal and spectral indices at the reverse-forward stage are very close to the observed ones.
As shown in \citet{zhang03}, in ISM environment the radiation from early forward shock always increases with time, which can not account for the shallow decay. For the wind environment, as shown in \citet{zou05}, the spectral index is $-p/2$ and the temporal index is $-(p-2)/2$ for the observed frequency being greater than the other typical frequencies. These two indices are very close to the observed behavior of the shallow decay phase, if the electron power law distribution index $p$ is slightly greater than 2.
We show the detailed model in section \ref{main}, and discussion and conclusion in section \ref{conclusion}.

\section{Model}
\label{main}

In the standard model of gamma-ray bursts, after internal shocks which produce prompt gamma rays, the sub-shells will merge up to a single cold shell with isotropic equivalent kinetic energy $E_0$ and Lorentz factor $\eta$. This shell proceeds in the external medium almost in a constant speed inside the deceleration radius $R_d$, which is defined as $M(R_d) \equiv {M_0}/{\eta}$ (\citealt{rees92}), where $M(R)$ is the medium mass inside radius $R$ and $M_0=E_0/(\eta c^2)$ is the mass of the shell.  If the circum-burst environment is a stellar wind with density $n(R) = A R^{-2}$, where $A = 3\times 10^{35} A_{*} {\rm cm}^{-1}$ is the wind parameter (\citealt{dai98}), the deceleration radius is $R_d \simeq 1.8 \times 10^{16} \eta_2^{-2} E_{0,53} A_{*,-1}^{-1} {\rm cm}$, and the corresponding observed deceleration time is  $t_{\oplus,d} \simeq 290 (1+z) \eta_2^{-4} E_{0,53} A_{*,-1}^{-1} {\rm s}$ (where $z$ is the cosmological redshift), which is very close to the break time between the shallow decay phase to the normal decay phase. We take the conventional notation $Q = Q_k \times 10^k$ in this paper.

Following \citet{spn98}, we consider the external shock radiation before the deceleration time. The number density and energy density of the shocked circum-medium (called region 2) are
$  n_2 \simeq
    3.3 \times 10^8 (1+z)^2 \eta_2^{-3} A_* t_{\oplus}^{-2}$, and
$  e_2 \simeq
    3.3 \times 10^{10} (1+z)^2 \eta_2^{-2} A_* t_{\oplus}^{-2}$ respectively.
The magnetic field in the comoving frame is
$B = \sqrt{8\pi \epsilon_B e} \simeq
36 \,A_{*,-1}^{{{1}\over{2}}}\,(1+z)\,t_{\oplus,2}
 ^ {- 1 }\,\epsilon_{B,-1}^{{{1}\over{2}}}\,\eta_{2}^ {- 1 } \,{\rm Gauss}$,
where $\epsilon_B$ is the equipartition factor for the magnetic energy density.
The peak spectral power is
$P_{\nu,\max} = (1+z) \sigma_T m_e c^2 \eta B/ (3 q_e)
 \simeq 1.3 \times 10^{-18}\,A_{*,-1}^{{{1}\over{2}}}\,(1+z)
 ^2\,t_{\oplus,2}^ {- 1 }\,\epsilon_{B,-1}^{{{1}\over{2}}}
 {\, \rm erg \, Hz^{-1} \, s^{-1} }$,
 where $\sigma_T$ is the Thomson cross section, $q_e$ is the electron charge.
The peak observed flux density is then
$f_{\nu,\max} = {N_e P_{\nu,\max} }/{(4\pi D^2)}
 \simeq 2.4 \times 10^{-23}\,A_{*,-1}^{{{3}\over{2}}}\,
 D_{28}^ {- 2 }\,(1+z)\,\epsilon_{B,-1}^{{{1}\over{2}}}\,\eta_{2}^2
   {\, \rm erg \, cm^{-2} \, Hz^{-1} \, s^{-1}}$,
 where $N_e$ is the total number of emitting electrons, and $D$ is the luminosity distance.
The synchrotron cooling Lorentz factor is
$\gamma_c = 6 \pi m_e c/(\sigma_T B^2 t_{co})
\simeq 31 \,A_{*,-1}^ {- 1 }\,(1+z)^ {- 1 }\,t_{\oplus,2}\,
 \epsilon_{B,-1}^ {- 1 }\,\eta_{2}
$
(where $t_{co}$ is the comoving dynamical time scale).
This corresponds to the cooling frequency
$\nu_c = (1+z)^{-1} \frac{\eta \gamma_c^2 q_e B}{2 \pi m_e c}
 \simeq 9.4 \times 10^{12}\,A_{*,-1}^ {- {{3}\over{2}} }\,(1+z)
 ^ {- 2 }\,t_{\oplus,2}\,\epsilon_{B,-1}^ {- {{3}\over{2}} }\,
 \eta_{2}^2 {\, \rm Hz}$.
The typical Lorentz factor of the electrons is
$\gamma_m = \frac{p-2}{p-1} \frac{\epsilon_e e}{n_2 m_e c^2}
 \simeq 9.8 \times 10^3 \,\epsilon_{e,-0.5}\,\eta_{2}\,\zeta_{1/6}
$,
and the typical synchrotron frequency is
$\nu_{m} = (1+z)^{-1} \frac{\eta \gamma_m^2 q_e B}{2 \pi m_e c}
 \simeq 9.3 \times 10^{17}\,A_{*,-1}^{{{1}\over{2}}}\,
 t_{\oplus,2}^ {- 1 }\,\epsilon_{B,-1}^{{{1}\over{2}}}\,
 \epsilon_{e,-0.5}^2\,\eta_{2}^2\,\zeta_{1/6}^2 {\, \rm Hz}$,
where $\zeta = 3 \frac{p-2}{p-1}$, and $p$ is the index of power law distributed electrons.
We neglect the synchrotron self-absorption for the X-ray emission as they are certainly much less than the frequency of the X-rays. The frequency of the observed X-rays is $\nu \sim 10^{18}$Hz, which is greater than both $\nu_c$ and $\nu_m$. Therefore, the flux density for the X-ray is
\begin{equation} \fl
f_\nu = 7.1 \times 10^{-26}\,A_{*,-1}^{{{21}\over{20}}}\,
 D_{28}^ {- 2 }\,t_{\oplus,2}^ {- {{1}\over{10}} }\,\epsilon_{B,-1}^{
 {{1}\over{20}}}\,\epsilon_{e,-0.5}^{{{6}\over{5}}}\,\eta_{2}^{{{21
 }\over{5}}}\,\nu_{18}^ {- {{11}\over{10}} }\,\zeta_{1/6}^{{{6}\over{
 5}}}
{\, \rm erg\, cm^{-2} Hz^{-1} s^{-1}},
\label{eq:fnu}
\end{equation}
for $p=2.2$. The value and the temporal behavior of the flux density are both consistent with the observed feature of the shallow decay phase. Noticing the temporal index is $-(p-2)/2$, which depends on $p$, the diversity of the observed shallow decaying index (generally between -0.5 and 0) can be well satisfied by the choosing of parameter $p$. The value of flux density for each individual burst may fit by the parameters shown in the equation above: $A_*, \epsilon_{B}, \epsilon_{e}$ and $\eta$. And the decelerating time $t_{\oplus,d}$ is determined by $\eta, E_0$ and $A_*$. 

To apply the model to individual burst, we have the following information to use. 
The break time from shallow decay phase to a normal phase $t_{\oplus,b} = t_{\oplus,d} \simeq 290 (1+z) \eta_2^{-4} E_{0,53} A_{*,-1}^{-1} {\rm s}$; the temporal index of the shallow decay phase $\alpha_2 \simeq -(p-2)/2$ for $\nu_X > (\nu_m, \nu_c)$,  $1/2$ for $\nu_c < \nu_X <\nu_m$, or $-(p-1)/2$ for $\nu_m < \nu_X < \nu_c$, respectively; 
the temporal index after the shallow decay phase is $\alpha_3 \simeq -(3p-2)/4$ for $\nu_X > (\nu_m, \nu_c)$,  $-1/4$ for $\nu_c < \nu_X <\nu_m$, or $-(3p-1)/4$ for $\nu_m < \nu_X < \nu_c$, which is normal decay in wind environment \citep{dai98}; the spectral index of the flux density for both these two phases should be the same, which is $\beta \simeq -p/2$ for $\nu_X > (\nu_m, \nu_c)$, $-1/2$ for $\nu_c < \nu_X <\nu_m$, or $-(p-1)/2$ for $\nu_m < \nu_X < \nu_c$; the flux density at $t_{\oplus,b}$ is from equation (\ref{eq:fnu}) but depends on the value of $p$ and the order between $(\nu_X, \nu_m, \nu_c)$, of which the details can be found in \cite{zou05}.
We apply our model to a few GRBs in the following.

{\it GRB 050319:}  The redshift of this burst is $z=3.24$, corresponding to a luminosity distance $D \simeq 8.75 \times 10^{28}$cm with cosmological parameters $\Omega_M=0.27, \Omega_\Lambda = 0.73$ and $H_0 =71 {\rm km\, s^{-1}\,Mpc^{-1}}$. The break time is $t_{\oplus,b} \sim 2.6\times 10^4$ s \citep{Cusumano2006}, which should be equal to $290 (1+z) \eta_2^{-4} E_{0,53} A_{*,-1}^{-1} {\rm s}$. The temporal indices of the shallow decay and after the decay are $\alpha_2 \sim 0.54$ and $\alpha_3 \sim -1.14$ respectively, and the spectral indices of the flux density for the X-ray are $-0.69$ and $-0.8$ \citep{Cusumano2006}, which is marginally the same for both phases. We choose $p=2.2$, and  $\nu_m < \nu_X < \nu_c$, which can marginally fit all the three parameters above. The flux density at $\sim 2.6\times 10^4$ s is $f_\nu \sim 3\times 10^{-29} {\rm erg\,cm^{-2}\,Hz^{-1}\,s^{-1}}$. For $p=2.2$, and  $\nu_m < \nu_X < \nu_c$, instead of equation (\ref{eq:fnu}) the flux density is $2.3 \times 10^{-23}\,A_{*,-1}^{{{9}\over{5}}}\,
 D_{28}^ {- 2 }\,(1+z)\,t_{\oplus,2}^ {- {{3}\over{5}} }\,
 \epsilon_{B,-1}^{{{4}\over{5}}}\,\epsilon_{e,-0.5}^{{{6}\over{5}}}\,
 \eta_{2}^{{{16}\over{5}}}\,\nu_{18}^ {- {{3}\over{5}} }  {\rm erg\,cm^{-2}\,Hz^{-1}\,s^{-1}}$. It requires $A_{*,-1}^{{{9}\over{5}}}\,
  \epsilon_{B,-1}^{{{4}\over{5}}}\,\epsilon_{e,-0.5}^{{{6}\over{5}}}\,
 \eta_{2}^{{{16}\over{5}}} \simeq 2.8 \times 10^{-3}$. Also considering the requirement of the order of typical frequencies $\nu_m < \nu_X < \nu_c$, there is still much space for the parameter choosing, like $A_{*,-1} \sim 10, \epsilon_{B,-1} \sim  10^{-4}, \epsilon_{e,-0.5} \sim 1, \eta_2 \sim 0.4$ and $E_{0,53} \sim 5.3$.

{\it GRB 081008:} The redshift of GRB 081008 is $z=1.967$ \citep{Yuan2010}, corresponding to $D \simeq 4.76 \times 10^{28}$ cm. The temporal indices are $\alpha_2 \sim -0.96$ and $\alpha_3 \sim -1.78$ respectively, with the break time $t_{\oplus,b} \sim 1.59 \times 10^4$ s. The spectral index of the X-ray is $\sim -1.12$ \citep{Yuan2010}. Choosing $p=2.9$ and $\nu_m < \nu_X < \nu_c$ agrees well with the temporal indices and spectral index above, the corresponding flux density at $10^{18}$ Hz is $6.7 \times 10^{-21}\,
 \eta_{2}^{{{39}\over{10}}}\,\nu_{18}^ {- {{19}\over{20}} }\,D_{28}
 ^ {- 2 }\,A_{*,-1}^{{{79}\over{40}}}\,\varepsilon_{e,-0.5}
 ^{{{19}\over{10}}}\,(1+z)\,t_{\oplus}^ {- {{19}\over{20}} }\,
 \varepsilon_{B,-1}^{{{39}\over{40}}}$. 
The observed flux density at the break time is $f_\nu \sim 2 \times 10^{-27} {\rm erg\,cm^{-2}\,Hz^{-1}\,s^{-1}}$, considering both the value of break time and the order of $(\nu_m,  \nu_X, \nu_c)$, we find a set of parameter $A_{*,-1} \sim 10^{-2}, \epsilon_{B,-1} \sim  10^{-2}, \epsilon_{e,-0.5} \sim 1, \eta_2 \sim 1.16$ and $E_{0,53} \sim 0.36$, which can fit the observations well.

\section{Conclusion and Discussion}
\label{conclusion}

We suggest an alternative simple model to explain the shallow decay phase of the early X-ray afterglow, which is that the external shock before decelerating emits the shallow decaying X-rays, while it transfers to a normal afterglow after the deceleration. The break time between the shallow decay phase and the normal decay phase is the deceleration time. As the transferring is just dynamics, which does not change the spectrum, it is well consistent with the observations. We applied our model to GRBs 050319 and 081008, which have the shallow decay phase, and found it can be well explained.

However, for some bursts with early optical emission observations, like GRB 060714 (\citealt{krimm07,liang07}), it seems at the break time of the X-rays, there is no break for the optical emission. This phenomenon may defy the break time of the X-rays is the deceleration time. But the non-achromatic break is a general phenomenon (\citealt{liang07}), and the intrinsic mechanism is not clear yet (\citealt{panaitescu06,zhang06}). It is possible that the optical and X-ray emissions are arising from different origin or have different microphysical parameters (\citealt{panaitescu06}). Another possible exception of our model is that some bursts have a shallow decay with index $\sim 1$, like GRB 050318 (\citealt{perri05}), which requires $p\sim 4$ in our model. Here we take this kind of decay as a normal X-ray afterglow with different microphysics rather than a real shallow decay. Furthermore, our model has some flexibility: the index $p$ can vary, and the environment can be slightly different a wind, i.e., the number density of the environment $n\propto A\, R^{-k}$ with $k$ being slightly different 2, which can be used for the complexity of observations.

%The shallow decaying index is $-(p-2)/2$ and the spectral index is $-p/2$, respectively. This induces the difference of these two indices to be a certain value 1. But the observations do not show this kind of relation. The reason is the typical frequency $\nu_m$ is close to the observed frequency, which makes the spectral index to be in the range of $[-p/2,-1/2]$.

%{\noindent \bf Acknowledgements}\\
\ack
We thank the discussion with LEI Wei-Hua. This work was supported by the National Natural Science Foundation of China (grants 10703002 and 11003004).

\newcommand{\mesz}{M\'esz\'aros}

%\section*{References}
%\begin{thebibliography}{99}
\References
\bibitem[{Cusumano et al.}(2006)]{Cusumano2006} Cusumano, G., et al. 2006, ApJ, 639, 316

\bibitem[{Dai \& Lu}(1998)]{dai98}Dai, Z. G., \& Lu, T. 1998, MNRAS, 298, 87

%\bibitem[{Dai \& Lu}(2001)]{DaiLu2001}Dai, Z. G., \& Lu, T. 2001, MNRAS, 324, L11

\bibitem[{Fan \& Piran}(2006)]{fan06}Fan, Y., \& Piran, T. 2006, MNRAS, 369, 197

\bibitem[{Ghisellini et al.}(2007)]{gs07}Ghisellini, G., Ghirlanda, G., Nava, L., \& Firmani, C. 2007, ApJ, 658, L75

\bibitem[{Kong et al.}(2010)]{kong10}Kong, S. W., Wong, A. Y. L., Huang, Y. F., \& Cheng, K. S. 2010, MNRAS, 402. 409

\bibitem[{Krimm et al.}(2007)]{krimm07}Krimm, H. A., et al. 2007, ApJ, 665, 554

\bibitem[{Liang, Zhang \& Zhang}(2007)]{liang07}Liang, E. W., Zhang, B. B., \& Zhang, B. 2007, MNRAS, 670, 565

\bibitem[{Liang et al.}(2009)]{liang09}Liang, E. W., Lv, H. J., Hou, S. J., Zhang, B. B., \& Zhang B. 2009, ApJ, 707, 328

\bibitem[{Nousek et al.}(2006)]{nousek06}Nousek, J. A., et al., 2006, ApJ, 642, 389

\bibitem[{O'Brien et al.}(2006)]{obrien06}O'Brien, P. T., et al. 2006, ApJ, 647, 1213

\bibitem[{Panaitescu et al.}(2006)]{panaitescu06}Panaitescu, A. 2006, MNRAS, 369, 2059

\bibitem[{Perri et al.}(2005)]{perri05}Perri, M., et al. 2005, A\&A, 442, L1

\bibitem[{Rees \& \mesz}(1992)]{rees92}Rees, M. J., \& \mesz, P. 1992, MNRAS, 258, 41p

\bibitem[{Sari, Piran \& Narayan}(1998)]{spn98}Sari, R., Piran T., \& Narayan, R., 1998, ApJ, 497, L17

\bibitem[{Toma et al.}(2006)]{toma06}Toma, K., Ioka, K., Yamazaki, R., \& Nakamura, T. 2006, ApJ, 640, L139

\bibitem[{Wu et al.}(2006)]{wu06}Wu, X. F., et al. 2006, 36th COSPAR Sci. Ass. \#731(arXiv:astro-ph/0512555)

\bibitem[{Xu \& Huang}(2010)]{xu10}Xu, M., \& Huang, Y. F. 2010, A\&A, 523, 5

\bibitem[{Yamazaki}(2009)]{yamazaki09}Yamazaki, R. 2009, ApJ, 690, L118

\bibitem[{Yuan et al.}(2010)]{Yuan2010}Yuan, F., et al. 2010, ApJ, 711, 870

\bibitem[{Zhang, Kobayashi \& \mesz}(2003)]{zhang03}Zhang, B., Koabyashi, S., \& \mesz, P. 2003, ApJ, 595, 950

\bibitem[{Zhang et al.}(2005)]{zhang05}Zhang, B., et al. 2005, ApJ, 642, 354

\bibitem[{Zhang}(2006)]{zhang06}Zhang, B. 2006, Advances in Space Research, 40, 1186

\bibitem[{Zou, Wu \& Dai}(2005)]{zou05}Zou, Y. C., Wu, X. F., \& Dai, Z. G. 2005, MNRAS, 363, 93

\bibitem[{Zou \& Dai}(2006)]{zou06a}Zou, Y. C., \& Dai, Z. G. 2006, ChJAA, 6, 551

\bibitem[{Zou, Xu \& Dai}(2006)]{zou06b}Zou, Y. C., Dai, Z. G., \& Xu, D. 2006, ApJ, 646, 1098

%\end{thebibliography}
\endrefs

\end{document}